**Resonant Spectroscopy of II-VI Self-Assembled Quantum Dots:**

**Excited States and Exciton-LO Phonon Coupling**

T.A. Nguyen, S. Mackowski*, H.E. Jackson, and L. M. Smith

*Deptartment of Physics, University of Cincinnati, Cincinnati, OH 45221-0011, USA*

J. Wrobel, K. Fronc, G. Karczewski, and J. Kossut

*Institute of Physics, Polish Academy of Science, Warsaw, Poland*

M. Dobrowolska and J. K. Furdyna

*Department of Physics, University of Notre Dame, IN, United States*

W. Heiss

*Institut für Halbleiter- und Festkörperphysik, Johannes Kepler Universität Linz, Austria*

## ABSTRACT

*Using resonantly excited photoluminescence (PL) along with photoluminescence excitation (PLE) spectroscopies, we study the carrier excitation processes in CdTe/ZnTe and CdSe/ZnSe self-assembled quantum dots (QDs). PLE spectra of single CdTe QDs reflect two major mechanisms for carrier excitation: The first, associated with the presence of sharp and intense lines in the spectrum, is a direct excited state – ground state transition. The second, associated with the appearance of up to four much broader excitation lines, is a LO phonon-assisted absorption directly into the QD ground states. LO phonons with energies of both QDs and ZnTe barrier material are identified in the PLE spectra. Resonantly excited PL measurements for the QD ensemble as a function of excitation energy makes it possible to separate the contributions of these two mechanisms. We find that for CdTe QDs the distribution of excited states coupled to*

*the ground states reflects the energy distribution of the QD emission, but shifted up in energy by 100 meV. This large splitting between excited and ground states in CdTe QDs suggests strong spatial confinement. In contrast, the LO phonon-assisted absorption shows significant size selectivity. In the case of CdTe dots the exciton-LO phonon coupling is strongly enhanced for smaller-sized dots which have higher emission energies. In contrast, for CdSe QDs the strength of exciton-LO phonon coupling is nearly uniform over the whole ensemble – that is, the dot energy distribution determines the intensities of LO phonon replicas. We show that for CdTe QDs after annealing, that is after an increase in the average dot size, the exciton-LO phonon interaction reflects the dot energy distribution, as observed for CdSe QDs.*

* corresponding author: Sebastian Mackowski, electronic mail: seb@physics.uc.edu

**INTRODUCTION**

Extensive spectroscopic studies of excitons in semiconductor quantum dots (QDs) show that either strong or weak LO phonon coupling can be observed depending on the energy scales involved. This variability of exciton-LO phonon coupling is caused by the discreteness of the quasi zero-dimensional energy levels in QDs and the ability to tune the energy levels through size or chemical composition engineering. The strong coupling regime is obtained when the energy distance between electronic levels in a QD approximately matches the LO phonon energy. In this non-adiabatic case a new quasi-particle, the polaron, is formed due to the strong interaction between carriers (electron, hole, or exciton) and the phonon bath [1-3]. Experimentally, the formation of the polaron is observed as a Stokes-shifted line in the resonantly excited spectrum of QDs [4]. The energy difference between the positions of such a polaron line and the excitation laser corresponds to the polaron binding energy. As a quasi-particle, the polaron is characterized by excited states within the quantum dot which are spaced by the LO phonon energy from each other, as has been observed in photoluminescence excitation (PLE) experiments on shallow QDs [5]. These polarons can be stable even at room temperature, and are characterized by very long lifetimes [6].

On the other hand, when the ground state – excited state (GS-ES) energy difference becomes much larger than the LO phonon energy, the exciton-LO phonon coupling enters the weak coupling regime. Characteristic of this regime, an exciton-LO phonon complex is formed which appears as a broad emission line spaced by multiple LO phonon energies from the laser [4]. These spectral signatures of excitons coupled weakly to LO phonons have been observed for different QD material systems by means of resonantly excited PL or PLE [7-12]. In all cases, however, the intensity dependence of LO phonon replica on the excitation energy has reflected the shape of the non-resonantly excited PL spectrum. Since in most cases the non-resonant PL is associated with the ground state energy distribution within QD ensemble, such a result indicates, that all probed QDs interact with LO phonons with the same strength. Somewhat indirectly, the size dependent exciton-LO phonon coupling has been reported through comparing the Huang-Rhys factors obtained for several InAs QD samples with different ground state energies [12]. Nevertheless, no clear evidence of size influence on exciton-LO phonon coupling in a single QD ensemble has been reported so far. This paucity of experimental evidence is explained by the several theoretical studies [13], which indicate that in order to observe size dependent exciton – LO phonon coupling, extremely small QDs sizes are required. Indeed, in the case of QDs with sizes comparable or smaller than the exciton Bohr radius, the change of the exciton wave-function due to confinement should result in a change of a dipole moment of the exciton. This, in turn, would change the interaction between the QD-confined exciton and the LO phonon.

In the majority of experiments reported so far, dots with sizes larger than the exciton Bohr radius have been studied. For instance, the typical size of II-VI CdSe QDs, where the exciton Bohr radius is equal to 3 nm, ranges from slightly larger than this value [14] to even 15 nm in a diameter [15]. In these cases then, the leakage of the exciton wave-function into the barriers is not expected to change significantly for QD size distributions within an ensemble. Consequently, for large QDs no size dependence of the exciton-LO phonon coupling is observed.

In this work we analyze resonant PL and PLE measurements for two II-VI semiconductor self-assembled QD systems. Self-assembled CdTe/ZnTe QD with sizes between 2 nm and 4 nm, which are significantly smaller than the exciton Bohr radius in bulk CdTe (10 nm). On the other hand, self-assembled CdSe/ZnSe QDs with sizes between 8 nm and 10 nm are significantly larger than the Bohr radius. The results reported here show that both CdTe QD and CdSe QDs are in the weak exciton-LO phonon coupling regime, while strongly confining electrons and holes. From resonantly excited PL of CdSe QDs we find only weak dependence of the exciton-LO phonon coupling on the QD emission energy. In contrast, in the case of smaller CdTe QDs the exciton-LO phonon coupling is strongly enhanced for QDs emitting at higher energies, i.e. those with smaller sizes. However, by increasing the average CdTe QD size *in the same sample* by rapid thermal annealing, we find that the exciton-LO phonon coupling becomes insensitive to the emission energy, similar to what is observed in the CdSe QDs case.

The excitons confined to CdTe QDs and CdSe QDs are both in the weak exciton – LO phonon coupling regime because the ES – GS energy splittings are substantially larger (~100 meV) than frequencies of LO phonons in these materials (~ 28 meV, on average). As we discuss below, our findings agree with recent theoretical calculation of the size dependence of exciton-LO phonon interaction in polar QDs [13]. The increase of the coupling for smaller QDs has been ascribed to changes in the wave-function of the exciton confined in a QD.

**SAMPLES AND EXPERIMENTAL DETAILS**

The CdTe/ZnTe and CdSe/ZnSe QD samples were grown by molecular beam epitaxy on GaAs substrates. Self-assembled CdSe (CdTe) QDs were formed by depositing 2.6 (4) monolayers on ZnSe (ZnTe) surface. A 50 nm thick ZnSe (ZnTe) capping layer covered the dot layer. Further details of the sample growth and their optical characterization can be found elsewhere [16,17]. It is important to note that the sample set chosen enables us to study QDs with considerably different lateral sizes. Namely, CdTe QDs are very small, and, as estimated by transmission electron microscopy and magneto-photoluminescence measurements, their lateral size is of the order of 2-4 nm in diameter [18]. On the other hand, the average lateral size of CdSe QDs studied here is larger and it approaches 8-10 nm in a diameter. In order to extend the available sizes of CdTe QDs we annealed CdTe QDs to increase their average size [19]. The annealing was performed in argon atmosphere at T= 470C for 15 seconds. From single dot spectroscopy in a magnetic field we estimate the average size of the CdTe QDs after annealing to be approximately 6-8 nm [20]. We note that for both II-VI QD structures studied here, no spectral evidence (PL or PLE) of a uniform wetting layer is observed, in contrast to the III-V semiconductor QDs [21].

Optical spectroscopy measurements were performed in a continuous-flow helium cryostat with sample temperatures of 6 K. Photoluminescence excitation (PLE) measurements of single CdTe QDs were carried out through an aperture in an opaque metal mask. The emission was excited by an Ar ion pumped dye laser (Rhodamine 590) focused to a spot diameter of 1.7 $\mu$m using a microscope objective. The QDs emission was dispersed by a DILOR XY triple spectrometer working in a subtractive mode and was detected by an $LN_2$-cooled CCD detector. Multichannel detection used in this experiment allows studying a number of QDs simultaneously. The excitation spectrum for a single QD can be easily obtained by analyzing the intensity of the particular emission line as a function of laser energy [22]. PLE spectra were normalized to the laser power. In the case of single dot PLE experiments, the spectral resolution is that of the dye laser, which was about 100 $\mu$eV. In order to obtain information about exciton – LO phonon coupling in our QDs, we performed resonantly excited PL measurements on CdSe QDs and on both as-grown and annealed CdTe QDs. In this case the emission was excited resonantly by a tunable dye laser using Rhodamine 590 and Coumarine 6 dyes. The spectral resolution for PL measurements was 70 $\mu$eV.

**EXPERIMENTAL RESULTS**

**PHOTOLUMINESCENCE EXCITATION OF SINGLE CdTe QUANTUM DOT**

In Fig. 1 we present typical PLE spectra obtained for two single CdTe QDs collected through an aperture with a diameter of 0.8 μm. The linewidths of these single dot PL emissions are around 100 μeV, typical for II-VI semiconductor QDs [23]. We have chosen two QDs to emphasize two distinct spectral features observed in the PLE spectrum. The broad excitation peak observed for QD1 (Fig. 1a) at a relative energy of 22 meV above the detection energy is attributed to LO phonon – assisted absorption into the QD ground state, as schematically shown in the inset of Fig. 1a. Similar LO phonon-related absorption lines have been recently found for single InAs QDs [24] and single CdSe QDs [25]. Typically we observe up to three such resonances and the linewidth of all replicas is approximately the same. The shape of these resonances can be deconvoluted into at least three Gaussian lines, yielding direct evidence that LO phonons with different energies which participate in this process [7,10]. In particular, the characteristic energies for bulk CdTe and ZnTe LO phonons are 21 meV and 24 meV, which agree with energy spacing observed for the resonance in Fig. 1a. The presence of ZnTe LO-phonon assisted absorption line suggests that the wave-function of the exciton confined in a QD is leaking out into the ZnTe barriers. We also observe a tail at energies less than 10 meV from the QD1 detection energy. This tail may originate from absorption mediated by acoustic phonons [25], but the possibility that at least part of this intensity is related to the stray laser light cannot be excluded.

Apart from excitations related to phonon-assisted processes we also observe a number of very intense and relatively sharp (with a linewidth less than 2 meV) excitation lines on a nearly background-free PLE spectrum, as shown for QD2 in Fig. 1b. From the measurements of tens of single CdTe QDs we find these lines predominantly in the range of 80-140 meV above the QD detection energy. The energy difference between these strong resonances and the detection energy varies from dot to dot revealing the influence of both size and chemical composition/strain inhomogeneities in the QD ensemble. We tentatively assign them to the direct excited state (ES) – ground state (GS) excitations (see inset to Fig.1b). It seems that on average the ES energy for CdTe QDs is significantly larger than, for instance, that observed for InAs [24] or CdSe QDs [25]. Since the ES-GS splitting is associated with the strength of spatial confinement, we may conclude that excitons in these CdTe QDs are very strongly confined. This finding agrees with a small lateral QD size (~ 2-4 nm in a diameter) estimated from previous magneto-PL spectroscopy results [18].

A detailed analysis of single CdTe QDs PLE results will be presented elsewhere. For the purpose of this paper we point out the presence of two dominant excitation mechanisms in CdTe QDs: direct ES-GS excitation (at around 100 meV) and LO phonon-assisted absorption.

**RESONANT PHOTOLUMINESCENCE OF QUANTUM DOTS**

The PLE data presented above shows that there are two dominant excitation processes in these structures: direct excitation into the QD ground states through LO phonon-assisted absorption or excitation into excited states followed by relaxation to ground states. This suggests that both excitation processes should be observed when resonantly exciting the QD ensemble. Indeed, as we shall show next, comparison of resonant PL of large QD ensembles with single-dot PLE makes it possible to draw more general conclusions about the importance of both QD excitation processes.

In Fig. 2a we show resonantly excited PL spectra measured through a 1.5 μm diameter aperture for CdTe QDs at T=6 K. The spectra are shifted vertically for clarity and compared with non-resonantly excited QD PL (shaded area in Fig. 2a). Solid triangles mark the laser energy for each spectrum. Depending on the excitation energy, the resonantly excited PL emission is composed of up to four clearly separated lines superimposed on a broad background. The energy spacing between these lines is roughly equal to 22 meV, which matches the energy spacing between detection energy and LO phonon absorption line observed in PLE spectrum. Moreover, the shape of the first LO phonon replica is very similar to the absorption line observed in Fig. 1a in the PLE experiments. We attribute these lines to the recombination of QDs excited directly into their ground states through LO phonon-assisted absorption. As one can see, the intensity of the LO phonons replicas depends strongly on the excitation energy. For instance, in the case of spectrum (2) in Fig. 2a, the highest emission intensity is associated with the third LO phonon replica, while spectrum number (4) in Fig. 2a is dominated by the first LO phonon emission. Moreover, the spectral features related to LO phonon enhanced PL become broader for the second and following replicas compared to the first replica. This broadening may be attributed to acoustic phonon scattering, which could be as efficient as higher order LO phonon relaxation processes [9].

We also performed resonant PL measurements for the larger-sized CdSe QDs in order to study the impact of QDs size on the excitation processes. From the PLE results reported for single CdSe QDs with similar sizes as CdSe QDs discussed here [25], in these structures both LO phonon-assisted absorption and direct ES – GS excitations are present. In Fig. 2b we show an example of PL spectra of CdSe QDs taken for different excitation energies (as marked with solid triangles). The shaded area represents the non-resonantly excited PL emission. As can be seen, resonantly excited PL spectra of CdSe QDs look qualitatively similar to the results found for CdTe QDs. All of the spectra consist of LO phonon assisted absorption lines and the broad background luminescence.

In order to determine the excitation energy dependence of the intensity of each component of the spectra we fit the data similar to that displayed in Fig. 2a and Fig. 2b with a series of Gaussian lineshapes. Extrapolating the results of single QD PLE over the whole ensemble, we assume that only the two aforementioned excitation processes determine the shape and the intensity of the resonant PL spectrum. We neglect, for instance, the effect of acoustic phonon-assisted hopping between QDs with different energies. Since all measurements were carried out at low temperature, this process should not substantially influence our analysis. In addition, the low excitation power used in these experiments allows us to rule out the possibility of Auger-type excitations. We also assume that for each particular excitation energy the ES-GS excitation process leads to a random QD occupation determined by the energy distribution of the GSs within the entire QD ensemble. As a consequence, we keep the energy and the linewidth of the background luminescence identical to the non-resonantly excited PL emission for all fits. Typical results of the fitting for CdTe QDs and CdSe QDs are presented in Fig. 3a and Fig. 3b, respectively. As can be seen, the fitted and the measured spectra agree reasonably well. By performing this fitting for all spectra, we are able to obtain the intensity of both the ES-GS transition and LO phonon-assisted absorption as a function of the excitation energy.

In Fig.4 we present the first LO phonon replica intensity dependence on the QD emission energy for CdSe QDs and both as-grown and annealed CdTe QDs. The energy dependence of the intensity of the first LO-phonon enhanced PL (solid points) is plotted along with the non-resonantly excited PL (shaded areas) measured for each sample. The solid lines are the non-resonant spectra shifted by 10 meV towards higher energies, while the dashed line is the least square fit to the data. The behavior of the second and third LO phonon replicas is similar to the one obtained for the first LO replica. However, the higher order replicas have somewhat broader linewidths probably due to scattering from acoustic phonons. We shall therefore concentrate here only on the emission originating from absorption with one LO phonon involved.

The results in Fig. 4 show that we observe quite different dependence of LO phonon intensity on the emission energy depending on the sample. In the case of the larger CdSe QDs the maximum of the intensity of the first LO phonon replica as well as its linewidth are very similar to those of the non-resonant PL emission. In contrast, for the as-grown CdTe QDs the highest intensity of the first LO phonon replica is observed at the energy shifted (by ~ 50 meV) towards higher energies with respect to the maximum of the non-resonant QD emission. Moreover, the linewidth of the LO phonon intensity dependence is narrower (~ 30 meV) than the inhomogeneous broadening of the CdTe QD PL (~ 70 meV). Interestingly, the same sample after annealing shows a behavior that is very similar to the one obtained for CdSe QDs. Since by annealing we increase the average size of CdTe QDs [19], this result strongly suggests that the exciton – LO phonon coupling in II-VI QDs depends sensitively on the size of QDs.

In the same way we have measured emission energy dependence of the LO phonon-assisted absorption process in QDs, we can also obtain the excitation energy dependence of the intensity of the emission which we associate with a direct ES-GS excitation in QDs. In Fig. 5 we plot the intensity of the background emission obtained for CdTe QDs and CdSe QDs at T=6 K. The results are compared with the non-resonant PL lineshapes (shaded areas in Fig. 5). In the case of CdTe QDs we find that the maximum enhancement of the ES-GS – related emission occurs for the laser energy approximately 100 meV above the maximum energy of the non-resonant PL. This energy agrees with the typical ES-GS splittings found in PLE spectra of single CdTe QDs.

In addition, the shape of the excitation energy dependence reflects the shape of the shifted non-resonant PL spectrum (solid line in Fig. 5). Importantly, we do not see, in agreement with PLE data, any indication of two-dimensional wetting layer for this sample. In fact, the intensity of ES-GS transition decreases rapidly after 100 meV. On the other hand, for CdSe QDs we observe a monotonous increase of the intensity ascribed to ES-GS excitation, even for the energies above 100 meV from the non-resonant PL spectrum (see Fig. 5). Possible explanations of this behavior are discussed later in the paper.

**DISCUSSION**

The results of resonant PL spectroscopy together with single-dot PLE measurements allow us to construct a consistent picture of excitation mechanisms in II-VI self-assembled QDs. In particular, we find two major processes that determine the optical properties of these structures. The first is associated with QD-confined exciton coupling to LO phonons while the other is related to direct transitions from excited to the ground state of the QD.

The ES-GS excitation process manifests itself by the presence of very sharp lines in the single dot PLE spectrum. The energy distance between these lines and the ground state QD energy varies from dot to dot, simply due to fluctuations of QD parameters within an ensemble. The most important factor which determines the value of GS - ES splitting, however, is the spatial confinement of excitons. For the very small CdTe QDs (the lateral size ~ 2-4 nm) the majority of ES – GS excitations are observed ~ 100 meV above the QD ground state. Interestingly, the linewidth of these resonances yields the lifetime of this transition in the picosecond range, which means that this excitation process is very effective.

The analysis of the resonant PL data is consistent with our interpretation of the PLE results. For CdTe QDs a background emission is observed, whose intensity depends strongly on the excitation energy. The energy dependence of this emission reflects to a good approximation the energy distribution of QDs within the ensemble. We conclude that the excited states in CdTe QDs have a similar energy distribution to the energy distribution of the QDs ground states, as has been measured also for other QDs systems [11,12,26]. Moreover, the energy difference between the maximum of the ES – GS transition and the non-resonant PL emission of ~ 100 meV corresponds well to the average energy spacing between excited and ground states obtained from PLE measurements.

Another explanation of the spectral features identified here as direct ES - GS transitions could be associated with the excitation of CdTe QDs through a two-dimensional wetting layer. However, both power dependent PL and PLE measurements show no evidence for such a continuum of states. Furthermore, if the background emission observed in resonant PL were originating from the WL, then the linewidth of its excitation energy dependence would be much narrower. As has been shown in previous studies [27], the PL linewidth of CdTe/ZnTe quantum wells does not exceed 40 meV. Furthermore, if the excitation were related to the two-dimensional wetting layer, it would feature the two-dimensional continuum density of states [7]. In contrast, the excitation energy dependence of the intensity of the ES-GS emission shows a well-resolved peak, so that the presence of a wetting layer can be ruled out.

In the case of CdSe QDs the intensity associated with ES-GS excitation shows a continuous increase with increasing excitation energy up to more than 100 meV above the maximum of non-resonantly excited PL. At this point we cannot exclude the presence of the wetting layer in this sample, but power dependent PL measurements certainly do not support the existence of such a wetting layer. Moreover, the fact that we do not observe any other spectral lines below 100 meV would indicate that the enhancement of the background emission intensity originate, as in CdTe QDs, from the ES-GS transition.

All of these observations indicate then that both sharp lines in micro-PLE spectra and the background emission in resonantly excited PL experiments are related to direct relaxation from an excited state down to the QD ground state. In both CdTe QDs and CdSe QDs the energy difference between ESs and GSs is significantly larger than the LO phonon energies characteristic for these materials. This justifies the assumption of a weak exciton-LO phonon coupling regime in these QDs.

Although the excitation energy dependence of the ES – GS excitation is qualitatively similar for both CdTe and CdSe QDs, the LO phonon-assisted processes show some important differences. In the case of the larger CdSe QDs, the maximum of the first LO phonon intensity dependence on the PL energy corresponds approximately to the maximum energy of non-resonantly excited PL emission (see Fig. 4). The linewidths of both these curves are also very similar. We argue that the LO phonon scattering strength in CdSe QDs is nearly independent of the QD size, and thus the intensity of the LO-phonon replica reflects only the ground state energy distribution within the QD ensemble. The behavior found for exciton – LO phonon coupling in CdSe QDs is similar to the direct ES – GS relaxation processes discussed earlier. We conclude that for CdSe QDs the size range and the size distribution in the CdSe QDs ensemble is too narrow to induce any significant changes in the overlap between electron and hole wave-function. This conclusion agrees reasonably well with the calculations by Melnikov and Beall Fowler [13] who show that indeed the exciton-LO phonon coupling is not very sensitive for the QD size studied here.

A situation where changing the overlap between electron and hole confined to QDs changes the strength of the exciton-LO phonon coupling, is in fact observed for the CdTe QDs, which are considerably smaller than CdSe QDs. As shown in Fig. 4, the CdTe QDs data feature a completely different dependence of the intensity of the first LO phonon replica on the excitation energy. Namely, the experimental results show that excitons in a narrow energy distribution of QDs couple much more strongly to LO phonons. First of all, the linewidth of the excitation energy dependence of the first LO phonon replica is much narrower than the QD energy distribution as a whole (revealed by non-resonant PL emission). Secondly, we observe a large energy separation between the non-resonant PL maximum and that of the first LO phonon enhancement. Since the intensity of the LO phonon replica depends on both exciton-LO phonon coupling and the density of states, it is necessary to deconvolute the dependence found for the first LO replica for CdTe QDs with the non-resonant spectrum. After the deconvolution we obtain the QD emission energy dependence of the LO phonon coupling. We find a strong increase of the LO phonon coupling for the dots with higher emission energies, probably reflecting their smaller size. Such dependence has been previously predicted theoretically for the similar QD systems [13]. Strong evidence that the effect observed for the as-grown CdTe QDs is related to size sensitivity of the exciton LO phonon coupling comes from the results obtained for the annealed CdTe QD sample. As shown previously [19,20], annealing of CdTe QDs leads to an increase of the average QD size and simultaneously improves the homogeneity of the ensemble. In such a case one would expect that the exciton-LO phonon coupling would approach the situation observed for CdSe QDs. Specifically, since the dots are larger and the size distribution is smaller, the experimentally measured intensity of the LO phonon replica would be determined mainly by the ground state energy distribution. The result presented in Fig. 4 for the CdTe QDs annealed at T=470C confirms this expectation.

**CONCLUSIONS**

By using single-dot PLE together with resonant PL spectroscopy, we identify two major carrier excitation mechanisms in CdTe QDs and CdSe QDs. The first one is direct excitation into the QD excited states followed by relaxation to the ground state and is predominantly governed by the ground state energy distribution within the QD ensemble. The second one is associated with direct excitation into the QD ground states through LO phonon-assisted absorption. By analyzing PL spectra obtained under resonant excitation, we show that the strength of exciton – LO phonon coupling increases significantly for QDs with lateral sizes smaller than the exciton Bohr radius, e.g. as-grown CdTe QDs. In contrast, for larger QDs (CdSe QDs) the strength of the exciton - LO phonon coupling is largely independent of the QD emission energy, and thus the intensities of LO phonon replicas are determined by the ground state energy distribution of the ensemble. For CdTe QDs with an increased average size obtained by rapid thermal annealing, we again find no dependence of the exciton-LO phonon coupling on the excitation energy. These results show that in the regime of weak exciton-LO phonon coupling the strength of this interaction may strongly increase for small QDs, in agreement with recent theoretical considerations.

**ACKNOWLEDGEMENT**

We gratefully acknowledge the financial support of the National Science Foundation through grants NSF DMR 0071797, 9975655 and 0072897 and the DARPA-SPINS Program. Work in Poland supported by PBZ-KBN-044/P03/2001.

**FIGURE CAPTIONS**

**Figure 1**. PLE spectra of two single CdTe QDs taken at T = 6K on 0.8 μm aperture. Both (a) LO phonon-assisted absorption and (b) excited state (ES) – ground state (GS) excitation are shown. The insets illustrate the respective energy diagrams of these excitation processes.

**Figure 2**. Resonantly excited PL spectra obtained for (a) CdTe QDs and (b) CdSe QDs at different laser energies (marked by the solid points). The spectra are shifted vertically. Also the non-resonantly excited PL emissions are shown.

**Figure 3**. Typical fits (black solid line) of the experimentally obtained PL spectra (solid points). The contribution of both excited state-ground state excitation (dashed line) and LO phonon-assisted absorption (grey lines) is shown for (a) CdTe QDs and (b) CdSe QDs.

**Figure 4**. The intensity of the first LO phonon replica obtained for as-grown CdTe QDs, annealed CdTe QDs and CdSe QDs as function of the emission energy. The shaded areas represent the non-resonantly excited PL spectra. Solid lines are the non-resonant PL spectra shifted by 10meV towards higher energies, while the dashed line is the least square fit to the data.

**Figure 5**. The intensity of the ES-GS excitation obtained for as-grown CdTe QDs and CdSe QDs as function of the excitation energy. The shaded areas represent the non-resonantly excited PL spectra. Solid line is the non-resonant PL spectrum shifted by 100meV towards higher energies, while the dashed line is the least square fit to the data. The ZnTe barrier energy is also shown.

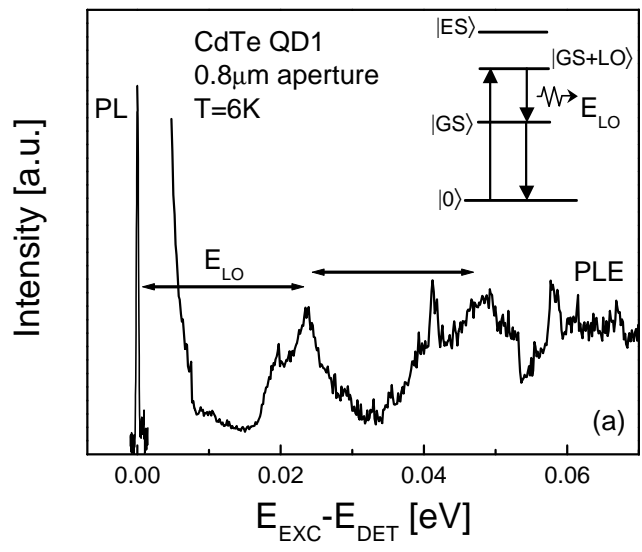

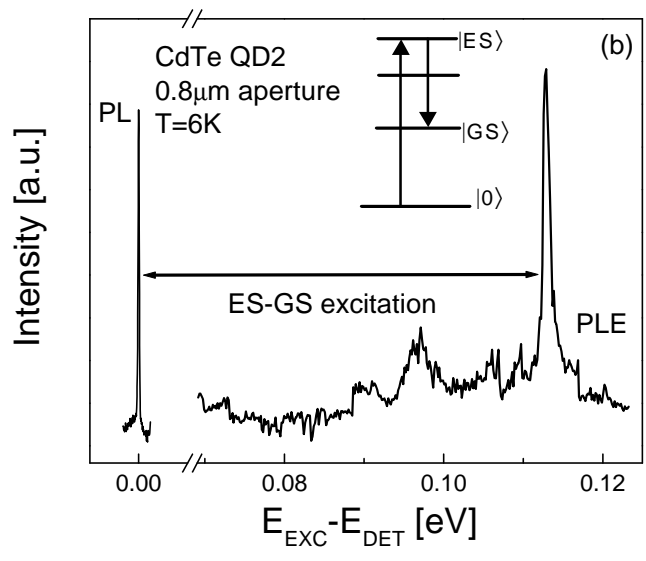

CdTe QD2
0.8μm aperture
T=6K

PL

ES-GS excitation

PLE

(b)

|ES⟩

|GS⟩

|0⟩

Intensity [a.u.]

$E_{EXC} - E_{DET}$ [eV]

0.00    0.08    0.10    0.12

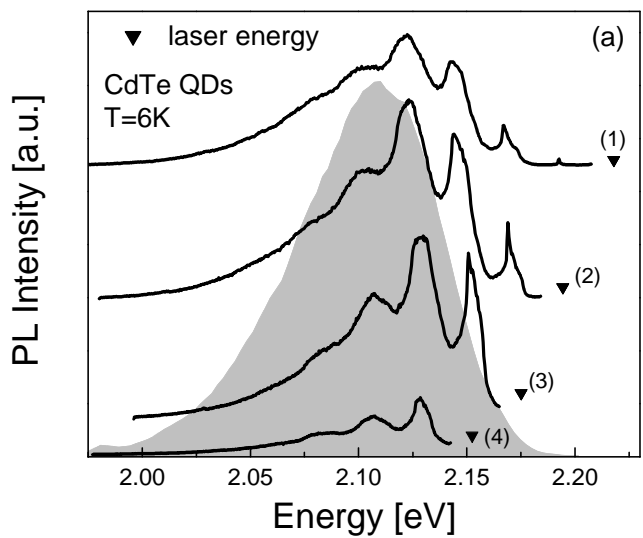

Fig.2a

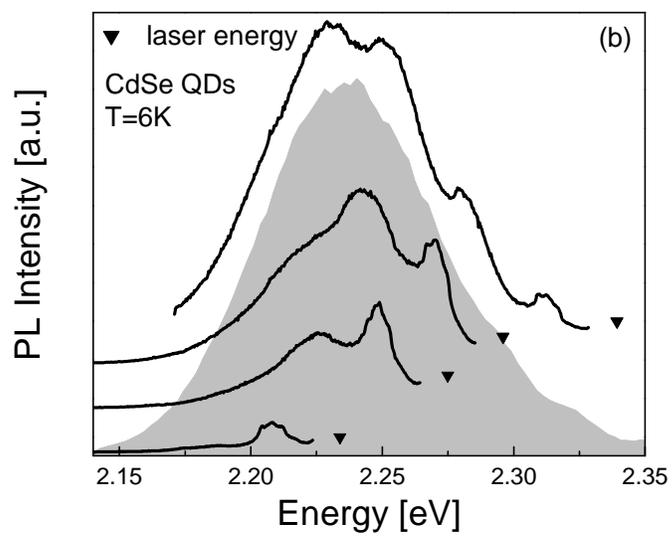

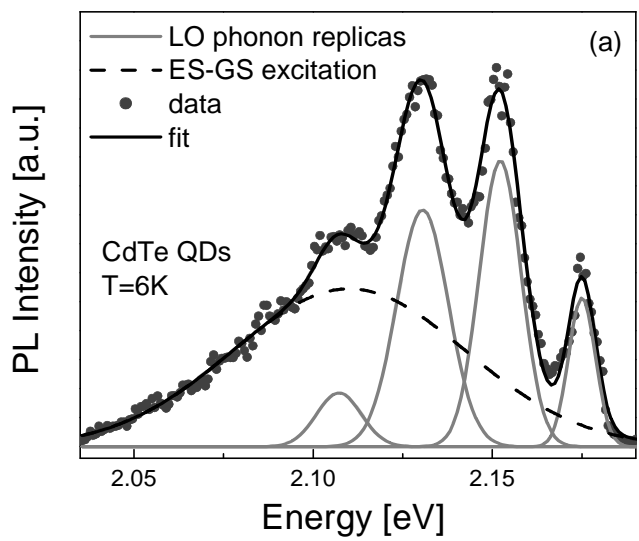

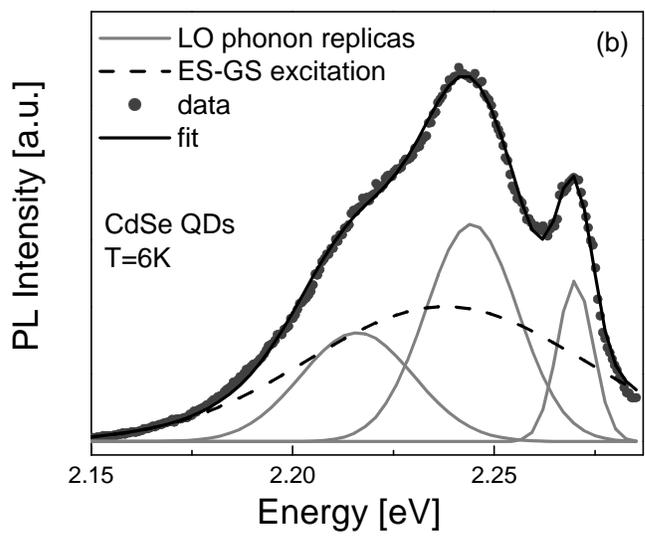

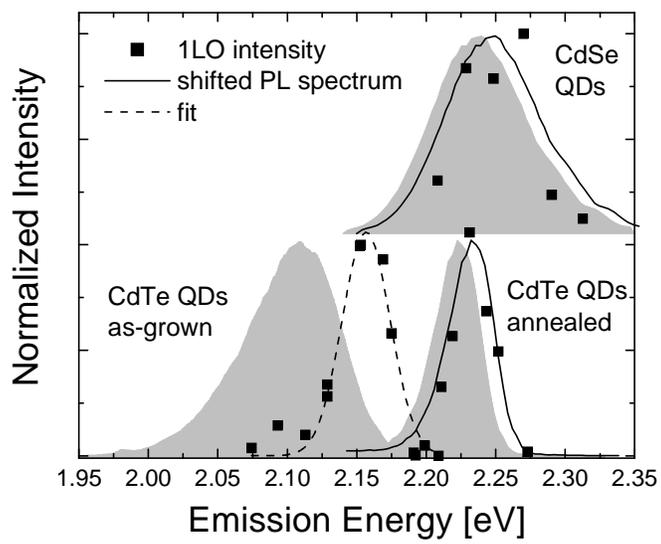

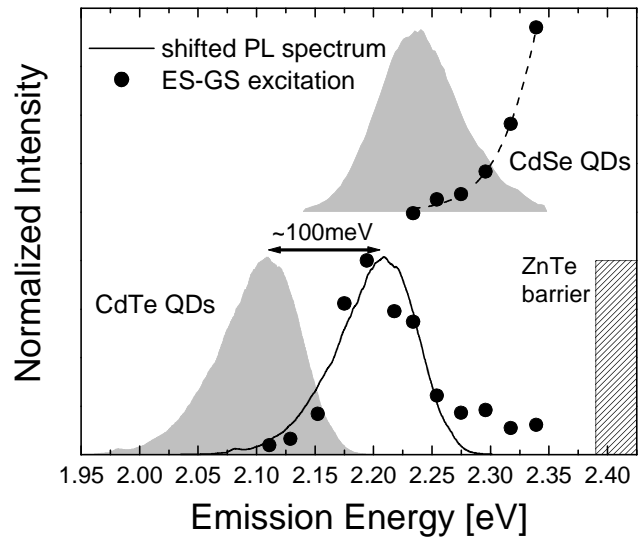